\documentclass[aps,prl,twocolumn,groupedaddress,amsmath,amssymb,showpacs]{revtex4} 

\usepackage{graphicx}% Include figure files 
\usepackage{dcolumn}% Align table columns on decimal point 
\usepackage{bm}% bold math 

\begin{document}

\title{Observation of Dirac Holes and Electrons in a Topological Insulator}

\author{A. A. Taskin}
\author{Zhi Ren}
\author{Satoshi Sasaki}
\author{Kouji Segawa}
\author{Yoichi Ando}

\affiliation{Institute of Scientific and Industrial Research,
Osaka University, Ibaraki, Osaka 567-0047, Japan}

\begin{abstract}

We show that in the new topological-insulator compound
Bi$_{1.5}$Sb$_{0.5}$Te$_{1.7}$Se$_{1.3}$ one can achieve a
surfaced-dominated transport where the surface channel contributes up to
70\% of the total conductance. Furthermore, it was found that in this
material the transport properties sharply reflect the time dependence of the
surface chemical potential, presenting a sign change in the Hall
coefficient with time. We demonstrate that such an evolution makes us
observe both Dirac holes and electrons on the surface, which allows us to
reconstruct the surface band dispersion across the Dirac point.

\end{abstract}


% 73.25.+i 	Surface conductivity and carrier phenomena
% 71.18.+y 	Fermi surface: calculations and measurements; effective mass,
%                     g factor
% 73.20.At 	Electron states at surfaces and interfaces - Surface states, 
%                 band structure, electron density of states
% 72.20.My      Conductivity phenomena in semiconductors and insulators - 
%                 Galvanomagnetic and other magnetotransport effects 

\maketitle

The three-dimensional (3D) topological insulator (TI) hosts a metallic
surface state that emerges due to a non-trivial Z$_2$ topology of the
bulk valence band \cite{FK,MB}. This peculiar surface state offers a new
playground for studying the physics of quasiparticles with unusual
dispersions, such as Dirac or Majorana fermions \cite{HK,QZ}. However,
most of the known TI materials are poor insulators in their bulk,
hindering transport studies of the expected novel surface properties
\cite{HK,QZ}. Last year, we discovered that the TI material
Bi$_{2}$Te$_{2}$Se (BTS) presents a high resistivity exceeding 1
$\Omega$cm \cite{BTS_Rapid}, which made it possible to clarify both the
surface and bulk transport channels in detail. Also, we found that in
our BTS sample the surface channel accounts for $\sim$6\% of the total
conductance. For this BTS compounds, Xiong {\it et al.} recently
reported a higher resistivity in the range of 5--6 $\Omega$cm, together
with pronounced surface quantum oscillations which possibly signify
fractional-filling states \cite{Xiong}. Since the source of the residual
bulk carriers in BTS is the acceptor states \cite{BTS_Rapid}, reducing
the number of anti-site defects working as acceptors in this promising 
material is an important challenge for the advancement of the TI research.

In this work, we tried to optimize the composition of BTS by reducing the
Te/Se ratio and introducing some Sb into Bi positions \cite{note_idea}, while
keeping the ordering of the chalcogen layers 
as in BTS (Fig. 1(a),  \cite{Nakajima}). The X-ray
powder diffraction patterns shown in Fig. 1(b) demonstrate that the
chalcogen ordering is still present in
Bi$_{1.5}$Sb$_{0.5}$Te$_{1.7}$Se$_{1.3}$ (BSTS), and we focus on this compound
in this Letter. We found that in BSTS one can achieve an
even larger surface contribution (up to 70\%) than in BTS. We also found
that the surface state of BSTS changes with time, and, intriguingly, we
observed quantum oscillations from Dirac holes, the negative
energy state of the Dirac fermions, as well as those from Dirac
electrons in the same sample at different time points. We show that this
time evolution can be used to reconstruct the surface band structure
across the Dirac point, providing a unique way to determine the 
dispersion relation of the surface state.

Single crystals of BSTS were grown by melting high purity (6N) elements
of Bi, Sb, Te, and Se with a molar ratio of 1.5:0.5:1.7:1.3 at
850$^{\circ}$C for 48 h in evacuated quartz tubes, followed by cooling
to room temperature over one week. For transport measurements, crystals
were cut along the principal axes, and cleaved just before the
measurements. Ohmic contacts were prepared by using
room-temperature cured silver paste. The resistivity $\rho_{xx}$ and the
Hall resistivity $\rho_{yx}$ were measured simultaneously by a standard
six-probe method \cite{note1} by sweeping the magnetic field $B$ between
$\pm$14 T at 0.3 T/min while stabilizing the temperature $T$ to within
$\pm$5 mK.

\begin{figure}\includegraphics*[width=8.5cm]{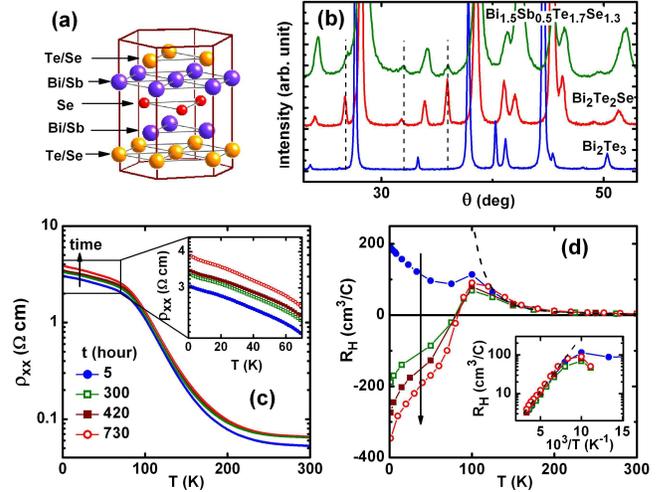}
\caption{(Color online) 
(a) Basic structure unit of Bi$_{1.5}$Sb$_{0.5}$Te$_{1.7}$Se$_{1.3}$ (BSTS).
(b) Comparison of the X-ray powder diffraction patterns of
BSTS, Bi$_{2}$Te$_{2}$Se, 
and Bi$_{2}$Te$_{3}$. Dashed lines indicate the positions of the peaks
characteristic of the ordering of Se and Te (Te/Se) layers.
(c) Temperature dependence of $\rho_{xx}$ measured repeatedly in time
in a cleaved 30-$\mu$m-thick BSTS sample. 
(d) Temperature dependence of the low-field $R_{H}$, presenting a sign change
with time. Dashed line represents the Arrhenius-law fitting to the data above 150 K, 
which is also shown in the inset.
}
\label{fig1}
\end{figure}

Freshly cleaved thin samples were used for studying the surface
transport in BSTS. As shown in Fig. 1(c), $\rho_{xx}$ in BSTS sharply
increases upon lowering temperature from 300 K, which is
characteristic of an insulator, but it saturates below $\sim$100 K due
to the metallic surface transport as well as the bulk impurity-band (IB)
transport \cite{BTS_Rapid}. This behavior is essentially the same as in
BTS. What is peculiar in BSTS is that $\rho_{xx}(T)$ increases slowly
with time and, furthermore, the low-temperature Hall coefficient $R_H$
changes sign with time in thin samples. As an example, Figs. 1(c) and
(d) show $\rho_{xx}(T)$ and $R_{H}(T)$ data of a 30-$\mu$m-thick
BSTS sample, measured repeatedly in about one month. In contrast to the
relatively small change in $\rho_{xx}$ [Fig. 1(c)], $R_H$ at
low-temperature exhibits rather drastic evolution [Fig. 1(d)] from
positive to negative, whereas $R_H$ at high temperature is essentially
unchanged with time. This suggests that the source of the time
dependence resides in the surface channel. In passing, $R_{H}(T)$ above
150 K is positive and demonstrates an activated behavior [shown by the
dashed line in Fig. 1(d)], signifying the thermal excitation of holes
into the valence band with an effective activation energy of about 60
meV. This is similar to what we observed in BTS \cite{BTS_Rapid}.

\begin{figure}\includegraphics*[width=8.5cm]{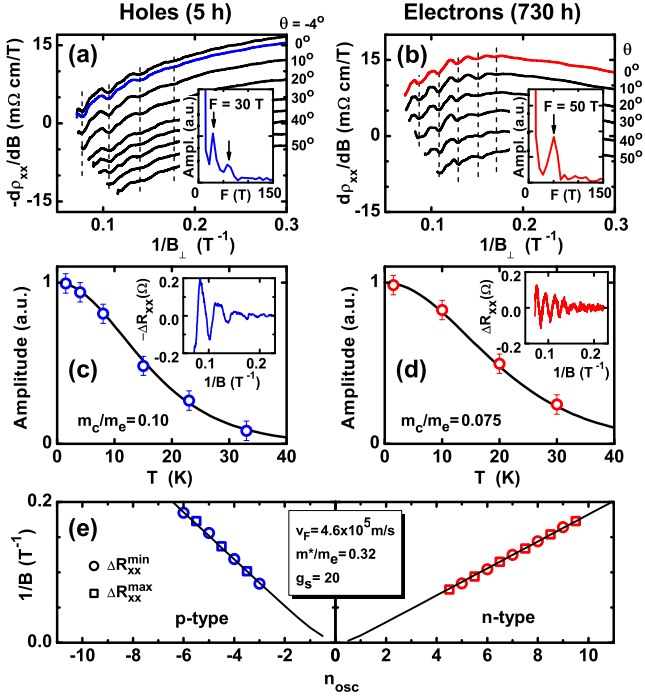}
\caption{(Color online) 
(a,b) $d\rho_{xx}/dB$ vs 1/$B_{\bot}$ measured in 
tilted magnetic fields in the $p$ state (5 h) and the $n$ state (730 h). 
All curves are shifted for clarity. 
Insets show the Fourier transforms of the data at $\theta$=0$^{\circ}$.
(c,d) Temperature dependences of SdH amplitudes
for $\theta$=0$^{\circ}$ shown in (a) and (b) and their theoretical fittings.
Insets show $\Delta R_{xx}$ vs. 1/$B$ after subtracting a smooth background.
(e) Landau-level fan diagram obtained from the oscillations in $\Delta R_{xx}$
measured at $T$ = 1.5 K and $\theta$ = 0$^{\circ}$; 
minima and maxima in $\Delta R_{xx}$ correspond to 
$n_{\rm osc}$ and $n_{\rm osc}+\frac{1}{2}$, respectively. 
} 
\label{fig2}
\end{figure}

To understand the nature of the time evolution, the Shubnikov-de Haas
(SdH) oscillations were measured in the aforementioned 30-$\mu$m-thick
sample twice, 5 h after cleavage (called $p$ state) and 725 h later
(called $n$ state), between which $R_H$ changed sign (the sample was
kept at 300 K in air). In BSTS, the oscillations do not fade out even
after long exposure to ambient atmosphere, as opposed to other TI
materials like Sb-doped Bi$_{2}$Se$_{3}$ \cite{Fisher_nm}. The SdH
oscillations were clearly observed in $\rho_{xx}(B)$ in our samples, but
they were not clear in $\rho_{yx}(B)$, so the following SdH
analysis is restricted to $\rho_{xx}(B)$. Figures 2(a) and (b) show
$d\rho_{xx}/dB$ for both the $p$ and $n$ states, plotted as a function
of 1/$B_{\bot}$ ($\equiv$ 1/$B\cos\theta$), where $\theta$ is the angle
between $B$ and the $C_{3}$ axis. Several equidistant maxima and minima
are easily recognized, and importantly, the positions of maxima and
minima depend solely on $B_{\bot}$, which signifies a 2D character of
the observed oscillations. Insets show the Fourier transform of the
oscillations taken at $\theta$=0$^{\circ}$. Two frequencies are clearly
seen in the $p$ state, but the second one (60 T) is probably a harmonic
of the primary frequency $F$ = 30 T. On the other hand, only the primary
$F$ = 50 T is seen in the $n$ state. The averaged Fermi wave number
$k_F$ is obtained by using the Onsager relation $F = (\hbar c / 2\pi
e)\pi k_F^2$, resulting in $k_{F}$ = 3.0$\times$10$^{6}$ cm$^{-1}$ and
3.9$\times$10$^{6}$ cm$^{-1}$ for the $p$ and $n$ states, respectively.
This corresponds to the surface hole (electron) concentration of
7.2$\times$10$^{11}$ cm$^{-2}$ (1.2$\times$10$^{12}$ cm$^{-2}$) for a
spin-filtered surface state. Fitting the standard Lifshitz-Kosevich
theory \cite{Shoenberg1984} to the temperature dependences of the SdH
amplitudes [Figs. 2(c) and (d)] gives the cyclotron mass $m_{c}$ of
(0.10$\pm$0.01)$m_{e}$ for holes and (0.075$\pm$0.003)$m_{e}$ for
electrons ($m_{e}$ is the free electron mass). Also, from the 
$B$-dependence of the SdH amplitudes [insets of Figs. 2(c) and (d)] 
one can obtain the scattering time $\tau$ of 5.8$\times$10$^{-14}$ s
(4.2$\times$10$^{-14}$ s) for holes (electrons) through the Dingle
analysis.

From the measured values of $k_F$ and $m_c$, one obtains the effective
Fermi velocity $v_{F}^{*}$ ($\equiv \hbar k_{F}$/$m_{c}$) of
3.5$\times$10$^{5}$ m/s and 6.0$\times$10$^{5}$ m/s for holes and
electrons, respectively. Now we discuss that this difference between the
two $v_{F}^{*}$ bears crucial information regarding the Dirac cone: 
If the surface state consists of ideal Dirac fermions, the Fermi
velocity is independent of ${\bf k}$ and is particle-hole symmetric.
However, the energy dispersions of the surface states in Bi-based TI
materials generally deviate from the ideal Dirac cone, and it can be
described as \cite{DasSarma2010}
\begin{equation}
E(k) = E_{DP}+ v_{F} \hbar k + \frac{\hbar^{2}}{2m^{*}}k^{2},
\end{equation}
where $E_{DP}$ is the energy at the Dirac point (DP), $v_{F}$ is the  
Fermi velocity at the DP, and $m^{*}$ is an effective mass 
that accounts for the curvature in $E(k)$. The effective Fermi velocity 
$v_{F}^{*}$ reflects the local curvature in $E(k)$ and can be expressed as
$v_{F}^{*}(k) = (\partial E/\partial k)/\hbar = v_{F}+ \hbar k/m^{*}$.
The $p$ and $n$ states correspond to the situations where the Fermi
energy $E_F$ is below or above the DP, respectively, and the time
evolution of $R_H$ is a manifestation of the time-dependent change of
the surface chemical potential. By using the $k_{F}$ and $v_{F}^{*}$
values obtained for the $p$ and $n$ states, we can solve simultaneous
equations to obtain $v_{F}$ = 4.6$\times$10$^{5}$ m/s and $m^{*}$ =
0.32$m_{e}$. This mean that the time evolution of the transport
properties allowed us to do a ``spectroscopy" of the surface state to
determine its dispersion, from which we can further estimate the
position of $E_F$ to lie 80 meV below (140 meV above) the DP in the $p$
($n$) state. Finally, the mean free path $\ell_{s}$ = $v_{F}^{*} \tau$
is about 20 nm (40 nm) and the surface mobility $\mu_{s}^{\rm SdH}$ = ($e
\ell_{s}$)/($\hbar k_{F}$) is about 1.0$\times$10$^{3}$ cm$^{2}$/Vs 
(9.8$\times$10$^{2}$ cm$^{2}$/Vs) in the $p$ ($n$) state.

To infer the Dirac nature of the surface state from the SdH oscillations
\cite{Mikitik1999,TA_Berry}, Fig. 2(e) shows the Landau-level (LL) fan
diagram, which presents the values of $1/B$ at the $n_{\rm osc}$-th
minima in the $\rho_{xx}$ oscillations as a function of $n_{\rm osc}$
\cite{TA_Berry}. In the case of ideal Dirac particles, the diagram for
holes (electrons) intersects the $n_{\rm osc}$-axis at $-0.5$ (0.5),
reflecting the $\pi$ Berry phase \cite{Mikitik1999,TA_Berry,Kim2005}.
However, in recent SdH studies of TIs
\cite{BTS_Rapid,Xiong,Fisher_nm,Ong2010,Morpurgo,HgTe} exact $\pi$ Berry
phase has rarely been reported and this deviation has been a puzzle. The
Zeeman coupling of the spins to the magnetic field can be a source of
this discrepancy \cite{Fisher_nm}, and in addition, the deviation of
$E(k)$ from the ideal linear dispersion also causes the Berry phase to
shift from $\pi$ \cite{TA_Berry}. We therefore followed the scheme of
Ref. \cite{TA_Berry} to see if the LL fan diagram obtained for BSTS can
be understood by considering these additional factors: The solid lines
in Fig. 2(e) show the theoretical LLs for the non-ideal Dirac fermions
\cite{TA_Berry} with the band parameters already obtained ($v_{F}$ =
4.6$\times$10$^{5}$ m/s and $m^{*}$ = 0.32$m_{e}$) and a surface
$g$-factor $g_s$ = 20 (which is the sole fitting parameter). Those lines
agree reasonably well with the experimental data for both the $p$ and
$n$ states, supporting not only the Dirac nature of the observed surface
state but also the conjecture that the holes and electrons reside on the
same Dirac cone.

\begin{figure}\includegraphics*[width=8.5cm]{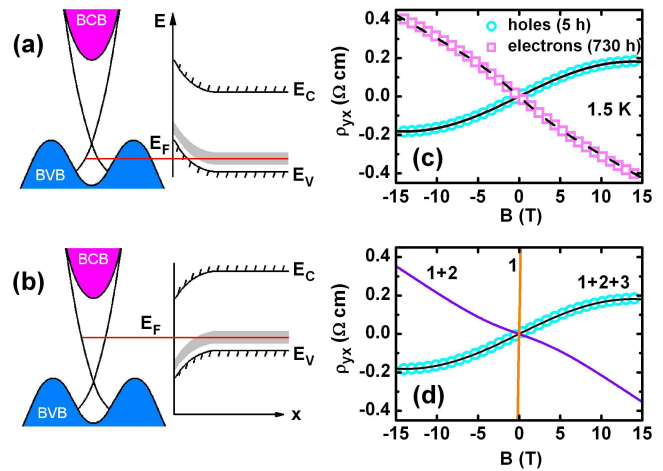}
\caption{(Color online) 
(a,b) Schematic picture of the bulk and surface states and the
surface band bending for the $p$ and $n$ states, respectively.
$E_{C}$ ($E_{V}$) is the energy of the conduction-band bottom
(valence-band top), and the shaded band depicts the impurity band. 
(c) $\rho_{yx}(B)$ data measured at 1.5 K in the $p$
and $n$ states. The dashed (solid) line is the
fitting of $\rho_{yx}(B)$ in the $n$-state ($p$-state) using the
two-band (three-band) model. (d) Decomposition of the three-band
$\rho_{yx}(B)$ fitting in the $p$ state (see text).
} 
\label{fig3}
\end{figure}

Now we discuss the mechanism for the time evolution of the transport
properties in BSTS. At low temperature, the Fermi level is pinned to the
IB in the bulk of the material \cite{BTS_Rapid}, so the observed
development of transport properties most likely comes from a change in
the surface as schematically shown in Figs. 3(a) and (b). To understand
the $p$ state where holes dominate the Hall response, one must assume
that an upward band-bending occurs just after cleavage, putting the
Fermi level below the DP and simultaneously creating a hole accumulation
layer (AL) near the surface [Fig. 3(a)]. The air exposure apparently causes
$n$-type doping on the surface as was reported for Bi$_2$Se$_3$
\cite{AnalytisPRB}, leading to a downward band bending [Fig. 3(b)].

In the above picture, there must be three transport channels in the $p$
state: surface Dirac holes, bulk IB, and the surface AL 
due to the band bending. Hence,
one may wonder if the SdH oscillations observed in the $p$ state might
actually be due to the AL, rather than the Dirac holes. Fortunately,
one can see that this is not the case, by analyzing the non-linear $B$
dependence of $\rho_{yx}$. In the following, we discuss the analyses of
the $\rho_{yx}(B)$ data, starting from the simpler case of the
$n$ state.

As in BTS \cite{BTS_Rapid}, the $\rho_{yx}(B)$ curves in the $n$ state
of BSTS can be well fitted with a simple two-band model described in
Ref. \cite{BTS_Rapid}. The dashed line in Fig. 3(c) is a result of such
fitting to the 1.5-K data, where the fitting parameters are the bulk
electron density $n_{b}$ = 2.3$\times$10$^{16}$ cm$^{-3}$, the bulk
mobility $\mu_{b}$ = 190 cm$^{2}$/Vs, and the surface mobility $\mu_{s}$
= 1250 cm$^{2}$/Vs (the surface electron density was fixed at
1.2$\times$10$^{12}$ cm$^{-2}$ from the SdH data). These parameters give
the residual bulk conductivity $\sigma_{b}$ of 0.73
$\Omega^{-1}$cm$^{-1}$, and the surface contribution to the total
conductance can be estimated as $G_{s}/(G_{s}+\sigma_{b}t) \approx$ 0.1,
where $G_{s} \approx$ 2.4$\times$10$^{-4}$ $\Omega^{-1}$ is the sheet
conductance of the surface and $t$ = 30 $\mu$m is the thickness.

In the $p$ state, the AL must also be taken into account, so we tried a
three-band analysis in which we assumed that the bulk carriers are the
same as in the $n$ state. The solid line in Fig. 3(c) shows a result of
the fitting to the 1.5-K data, where the fitting parameters are the AL
mobility $\mu_{s'}$ = 770 cm$^{2}$/Vs, the AL sheet conductance $G_{s'}$
= 2.2$\times$10$^{-3}$ $\Omega^{-1}$, and the Dirac-hole mobility
$\mu_{s}$ = 1170 cm$^{2}$/Vs (the Dirac hole density was fixed at
7.2$\times$10$^{11}$ cm$^{-2}$ from the SdH data). To understand the
relative importance of the three channels, it is instructive to examine
the individual contributions to the total $\rho_{yx}$: As shown in Fig.
3(d), the putative $\rho_{yx}(B)$ due solely to the surface Dirac holes
(curve 1) is strongly modified when the residual bulk contribution is
added (curve 1+2), but it is still very different from the measured
$\rho_{yx}(B)$; only when the third contribution of the AL is added
(curve 1+2+3), the $\rho_{yx}(B)$ behavior is satisfactorily reproduced.

Based on the above analysis, one can see that it is impossible to
interpret the SdH oscillations in the $p$ state to originate from the
AL: If the SdH oscillations were due to the AL, the third transport
channel must be the surface Dirac holes; however, the sheet carrier
density $n_{s'}$ of the third channel is estimated as $n_{s'}$ =
$G_{s'}/(e\mu_{s'}) \approx$ 1.8$\times$10$^{13}$ cm$^{-2}$, which is
too large for Dirac holes for which $k_F \lesssim$ 5$\times$10$^{6}$
cm$^{-1}$ \cite{ARPES} and hence $n_{s'}$ must be $\lesssim$ 2$\times$10$^{12}$
cm$^{-2}$. Therefore, one can safely conclude that the SdH oscillations
are due to the Dirac holes.

\begin{figure}\includegraphics*[width=6.2cm]{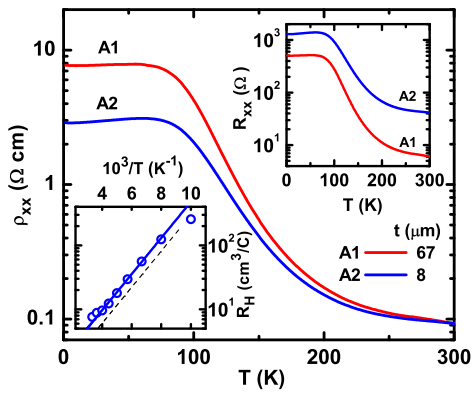}
\caption{(Color online) 
Temperature dependences of $\rho_{xx}$ in 0 T measured in a BSTS sample before
and after reducing its thickness.
Upper inset shows the plot of $R_{xx}(T)$ for the same set of data.
Lower inset shows the high-temperature activation behavior in $R_H(T)$ for
the 8-$\mu$m sample, which is compared with the similar behavior
observed in the 30-$\mu$m sample [Fig. 1(d) inset] shown as a dashed line.
} 
\label{fig4}
\end{figure}

Lastly, we show that the surface conductance in BSTS can be reasonably
estimated by simply changing the thickness. Figure 4 shows the
temperature dependences of $\rho_{xx}$ of a different sample, measured
first in 67-$\mu$m thickness, and later cleaved down to 8-$\mu$m thick
\cite{note2}. The overall behavior of $\rho_{xx}(T)$ is similar between
the two curves, but a striking difference lies in their low-temperature
saturation values $\rho^{\rm sat}_1$ (for 67 $\mu$m) and $\rho^{\rm
sat}_2$ (for 8 $\mu$m): it is lower in the thinner sample, implying a
larger relative surface contribution. Note that the {\it resistance}
$R_{xx}$ duly increases upon reducing the thickness, as shown in the
upper inset of Fig. 4. Instructively, the difference in $\rho_{xx}$
disappears at high temperature where the bulk conduction dominates. From
the observed difference in $\rho^{\rm sat}_{i}$ one can estimate the
surface and bulk contributions to the total conductivity by using
$\rho^{\rm sat}_{i} = [( G_{s}/t_{i}) + \sigma_{b}]^{-1}$. We obtain
$G_{s} \approx$ 1.8$\times$10$^{-4}$ $\Omega^{-1}$ and $\sigma_{b}
\approx$ 0.1 $\Omega^{-1}$cm$^{-1}$, and this $\sigma_{b}$ is much
smaller than that in the 30-$\mu$m-thick sample. This is probably
because the number of acceptors is smaller in this second sample
\cite{note3}, as can be inferred in the high-temperature $R_H$ behavior
[lower inset of Fig. 4]. The obtained values of $G_s$ and $\sigma_b$
allows us to calculate the fraction of the surface contribution to the
total conductance, $G_{s}/(G_{s}+\sigma_{b}t_i)$, for the 67- and
8-$\mu$m thick samples to be 0.2 and 0.7, respectively. Therefore, when
the thickness of a BSTS sample is reduced to $\lesssim$ 10 $\mu$m, one
can achieve a bulk TI crystal where the topological surface transport is
dominant over the bulk transport.

In summary, we show that one can achieve a surface-dominated transport
in the new TI compound Bi$_{1.5}$Sb$_{0.5}$Te$_{1.7}$Se$_{1.3}$. The
surface band bending and its time dependence makes it possible to
observe the SdH oscillations of both Dirac holes and electrons, with
which we could determine the surface band dispersion across the Dirac
point. In addition, by analyzing the non-linear $\rho_{yx}(B)$ curves,
we could identify the role of the surface accumulation layer in the 
transport properties. Obviously, this material offers a
well-characterized playground for studying the topological surface
state.

\begin{acknowledgments} 

This work was supported by JSPS (NEXT Program and KAKENHI 19674002), 
MEXT (Innovative Area ``Topological
Quantum Phenomena" KAKENHI), and AFOSR (AOARD 10-4103).

\end{acknowledgments}


\begin{thebibliography}{99}
\bibitem{note_idea}
Since the anti-site defects occur between Bi and Te, reducing
the Te/Se ratio would suppress the anti-site; at the same time, Se in
the outermost layer is prone to vacancy creations, which 
is suppressed by Sb doping.
\bibitem{note1}
The sample is suspended on gold wires in order to insure 
that both surfaces are in the same condition.
\bibitem{note2}
The cleaved 8-$\mu$m-thick sample showed a gradual increase 
in $\rho_{xx}$ similar to the 30-$\mu$m-thick sample,
so we only used relaxed samples (in which the AL is absent) for the 
thickness-dependence study.
\bibitem{note3}
Unfortunately the SdH oscillations were not observed in this sample,
probably due to a lower surface mobility.
\end{thebibliography}
\end{document}